\documentstyle[amstex,amssymb,preprint,aps]{revtex}

\begin{document}
\title{New chiral-symmetry-breaking operators in\\
pseudoscalar QCD sum rules\thanks{%
HD-TVP-01-21} }
\author{Hilmar Forkel}
\address{Institut f{\"u}r Theoretische Physik, Universit{\"a}t Heidelberg, \\
D-69120 Heidelberg, Germany}
\date{December 2001}
\maketitle

\begin{abstract}
Nonperturbative Wilson coefficients associated with the leading
chiral-symmetry-breaking operators in the operator product expansion of the
pseudoscalar QCD correlation function are derived. Implementation of the
new, instanton-induced operators enables the corresponding spectral sum rule
to reproduce the small pion mass scale, thereby reconciling it with
Goldstone's theorem. The same operators suppress the contributions of pionic
resonances. Several predictions and structural insights from the new sum
rule are discussed.
\end{abstract}


The arguably most characteristic pion property is its mass well below all
other hadronic mass scales. This special feature has long been understood as
a consequence of spontaneous chiral symmetry breaking (SCSB) in the QCD
vacuum \cite{pag78}, which renders the pion a (quasi-) Goldstone boson and
manifests itself in the finite vacuum expectation values of
chirality-changing operators like the quark condensate $\left\langle \bar{q}%
q\right\rangle $. One would therefore expect the QCD sum-rule approach \cite
{shi79}, which explicitly links hadron properties by means of the operator
product expansion (OPE) to such condensates, to provide a privileged source
of information on the Goldstone nature of the pion and, in particular, to
predict its small mass scale as a consequence of chiral-symmetry-breaking
(CSB) condensates.

Surprisingly, this expectation is not borne out by the existing analyses of
the pseudoscalar sum rule. They fail to predict a pion mass below standard
hadronic scales \cite{shi79}, and\ the contributions from chirality-changing
condensates are too strongly suppressed (by factors of the light quark
masses) to have a notable impact on the resulting pion properties. This
confronts us with a fundamental puzzle which threatens to unsettle the
conceptual basis of the QCD-sum-rule approach: why is the conventional OPE
of the pseudoscalar correlator practically blind to spontaneous chiral
symmetry breaking, i.e. to exactly that element of QCD vacuum physics which
profoundly affects the Goldstone boson channel?

Additional contributions - neglected in the standard OPE but potentially
enhanced in the pseudoscalar channel - could provide an attractive
resolution of this puzzle. One such contribution, a low-dimensional power
correction conjectured to originate from ultraviolet-sensitive physics, has
been proposed in \cite{che99}. Besides probably being small \cite{goe01},
however, this term is chirally invariant and therefore unlikely to
significantly improve the pion mass prediction. Hard (so-called direct)
instanton corrections to the unit-operator coefficient \cite
{ges80,nov81,shu83}, although of subtantial size, also proved unable to
generate the physical pion mass scale \cite{ste97}. The reason is probably
the same as above: these contributions are chirally invariant (as their
perturbative counterparts) and multiply the likewise chirally invariant unit
operator.

The above examples illustrate that information related to SCSB can enter the
OPE only through chiral-symmetry-breaking condensates (since it originates
from soft vacuum physics). In search for missing physics related to such
condensates, we will identify and calculate in this letter the leading
nonperturbative corrections to the Wilson coefficients associated with the
lowest-dimensional CSB operators. These new contributions arise from small
instantons and have the potential to resolve the above-mentioned puzzle
since instantons couple particularly strongly to pseudoscalar interpolating
fields \cite{ges80,nov81} and generate Wilson coefficients which are not
suppressed by light-quark mass factors (in contrast to their perturbative
counterparts).

\bigskip We start from the $\pi ^{0}$ correlation function 
\begin{equation}
\Pi \left( x\right) =\left\langle 0|T\,j_{\pi ^{0}}\left( x\right) j_{\pi
^{0}}\left( 0\right) |0\right\rangle ,
\end{equation}
based on the pseudoscalar interpolator 
\begin{equation}
j_{\pi ^{0}}=\sqrt{\frac{1}{2}}\left( \bar{u}i\gamma _{5}u-\bar{d}i\gamma
_{5}d\right) ,  \label{pi0curr}
\end{equation}
which has an instanton-improved OPE (IOPE) (for a review see \cite{for02})
of the general form \ 
\begin{eqnarray}
\Pi ^{\left( IOPE\right) }(Q^{2}) &=&i\int d^{4}x\,e^{iqx}\left\langle
0|T\,j_{\pi ^{0}}\left( x\right) j_{\pi ^{0}}\left( 0\right) |0\right\rangle
^{\left( IOPE\right) }  \nonumber \\
&=&\sum_{n}C_{\hat{O}_{n}}\left( Q^{2};\mu \right) \left\langle \hat{O}%
_{n}\left( \mu \right) \right\rangle  \label{genope}
\end{eqnarray}
($Q^{2}=-q^{2}\geq 1$ GeV, $\mu \lesssim 1$ GeV is the operator
renomalization scale).

The perturbative parts $C_{\hat{O}_{n}}^{\left( pert\right) }$ of the Wilson
coefficients, to $O\left( \alpha _{s}\right) $ for the unit operator and to
leading order for all remaining operators up to mass dimension $d=6$, are 
\cite{shi79,bec81} 
\begin{eqnarray}
C_{1}^{\left( pert\right) }\left( Q^{2};\mu \right) &=&\frac{3}{8\pi ^{2}}%
Q^{2}\ln \frac{Q^{2}}{\mu ^{2}}\left[ 1+\frac{17}{3}\frac{\alpha _{s}}{\pi }-%
\frac{\alpha _{s}}{\pi }\ln \frac{Q^{2}}{\mu ^{2}}\right] , \\
C_{\bar{q}q}^{\left( pert\right) }\left( Q^{2}\right) &=&-\frac{m_{q}}{Q^{2}}%
,\text{ \ \ \ \ \ }C_{\alpha G^{2}}^{\left( pert\right) }\left( Q^{2}\right)
=\frac{1}{8\pi Q^{2}},\text{ \ \ \ \ \ } \\
C_{\hat{O}_{4}}^{\left( pert\right) }\left( Q^{2}\right) &=&-C_{\hat{O}%
_{5}}^{\left( pert\right) }\left( Q^{2}\right) =-\frac{\pi }{Q^{4}},
\end{eqnarray}
where $\hat{O}_{4,5}$ are the four-quark operators 
\begin{align}
\hat{O}_{4}& =-\alpha \left( \bar{u}\sigma _{\mu \nu }t^{a}u\right) \left( 
\bar{d}\sigma _{\mu \nu }t^{a}d\right) , \\
\hat{O}_{5}& =\frac{1}{2}\alpha \left[ \left( \bar{u}\sigma _{\mu \nu
}t^{a}u\right) ^{2}+\left( \bar{d}\sigma _{\mu \nu }t^{a}d\right) ^{2}\right]
+\frac{1}{3}\alpha \left( \bar{u}\gamma _{\mu }t^{a}u+\bar{d}\gamma _{\mu
}t^{a}d\right) \left( \sum_{u,d,s}\bar{q}\gamma _{\mu }t^{a}q\right)
\end{align}
($t^{a}=\lambda ^{a}/2$, where $\lambda ^{a}$ are the Gell-Mann matrices).

Each term in the IOPE (\ref{genope}) factorizes into contributions from hard
modes with momenta $\left| k\right| >\mu $, contained in the $C_{\hat{O}%
_{n}} $, and from soft modes with $\left| k\right| \leq \mu $ in the
operators $\hat{O}_{n}$. Despite widespread belief (based on asymptotic
freedom) there are hadron channels where this does not even approximately
amount to a factorization of perturbative and nonperturbative physics (at $%
\mu \sim 1$ GeV). Indeed, early conjectures \cite{nov81} and substantial
recent evidence \cite{shu83,for93,for02} corroborate that (semi-) hard
nonperturbative contributions due to small instantons are quantitatively
relevant or even dominant in several hadronic correlators.

The pseudoscalar correlator, in particular, is known to receive
exceptionally strong direct-instanton contributions to the unit-operator
coefficient, 
\begin{equation}
C_{1}^{\left( I+\bar{I}\right) }\left( Q^{2}\right) =\int d\rho n\left( \rho
\right) \frac{\bar{m}_{u,2}^{2}\left( \rho \right) +\bar{m}_{d,2}^{2}\left(
\rho \right) }{\bar{m}_{u,2}^{2}\left( \rho \right) \bar{m}_{d,2}^{2}\left(
\rho \right) }\left( Q\rho \right) ^{2}K_{1}^{2}\left( Q\rho \right)
\label{cinst1}
\end{equation}
($I$, $\left( \bar{I}\right) $ refers to the (anti-) instanton closest to $x$%
, $n\left( \rho \right) $ is the vacuum distribution of instantons with size 
$\rho $, and $K_{1}\left( z\right) $ is a McDonald function \cite{abr65}),
which arise from the propagation of both quark and antiquark (ejected by (%
\ref{pi0curr})) in the zero mode of the instanton field\footnote{$%
C_{1}^{\left( I+\bar{I}\right) }\left( Q^{2}\right) $ has the same momentum
dependence as several instanton contributions (which arise from the
analogous diquark loop) to baryon sum rules \cite{for93,aw99}.} and can be
obtained by means of semiclassical techniques \cite{nov81,for02}. Due to
interactions with ambient, long-wavelength vacuum fields (including other
instantons) \cite{shi80} the quarks aquire an effective mass $\bar{m}%
_{q,2}\left( \rho \right) $ (the index indicates that two quarks are
propagating in zero modes). The quantitative sum-rule analysis below only
requires the value of $\bar{m}_{q,2}$ at the average instanton size $\bar{%
\rho}\simeq 0.33$ fm for which we adopt the recent estimate $\bar{m}%
_{q,2}\left( \bar{\rho}\right) \equiv \bar{m}_{q,2}\simeq 85$ MeV obtained
from instanton-liquid model (ILM) simulations of the pseudoscalar correlator 
\cite{fac01}. The $\bar{m}_{q,2}$-dependence of the results will be
discussed in \cite{for202}.

All so far considered contributions to the IOPE coefficients are either
associated with chirally-invariant operators or too strongly suppressed
(note the factor $m_{q}$ in $C_{\bar{q}q}^{\left( pert\right) }$) to
generate more than minute corrections to the pion mass. Hence at this stage
- which represents the current state of the art - the IOPE does contain
virtually no information on the soft vacuum fields which are responsible for
the spontaneous breakdown of chiral symmetry. Therefore it is not surprising
that the corresponding sum rule\ is unable to generate the low mass scale
which characterizes the Goldstone pion \cite{ste97}.

As we have argued above, there are reasons to believe that the missing
information on SCSB is activated by nonperturbative contributions to the
Wilson coefficients of chirally noninvariant operators which so far went
unnoticed. Furthermore, direct instantons are promising candidates for such
contributions since (i) they provide the leading nonperturbative deviations
from asymptotic freedom, (ii) they are likely to play an important r\^{o}le
in the dynamics of SCSB and in the strong flavor-mixing among pseudoscalar
mesons \cite{ges80}, (iii) their small average size $\bar{\rho}\lesssim \mu
^{-1}$ allows them to contribute strongly to the Wilson coefficients, (iv)
light-quark-mass suppression factors are absent, (v) the sensitivity of
spin-0 meson channels to instanton-induced short-distance physics is enhanced%
\footnote{%
The same holds for spin-0 glueballs \cite{for200}.} \cite{sch98}, and
finally (vi) recent lattice measurements find the pseudoscalar correlator
dominated by contributions from instanton-induced quark (quasi-) zero modes 
\cite{iva98}.

We are thus led to derive the instanton contributions to the Wilson
coefficients associated with the dominant (i.e. lowest-dimensional)
chiral-symmetry-breaking operators of the IOPE, $\bar{q}q$ and $g\bar{q}%
\sigma Gq$. They can be calculated as the leading terms in the semiclassical
expansion of the correlator around the (anti-) instanton in the background
of long-wavelength quark and gluon vacuum fields. We postpone a detailed
description of this calculation to \cite{for202} and present here just the
results, 
\begin{eqnarray}
C_{\bar{q}q}^{\left( I+\bar{I}\right) }\left( Q^{2}\right) &=&-\frac{\pi ^{2}%
}{2}\int d\rho n\left( \rho \right) \frac{\rho ^{4}}{\bar{m}_{q,1}\left(
\rho \right) }  \nonumber \\
&&\times \int_{0}^{\infty }\frac{d\alpha }{\alpha ^{2}}\,_{1}F_{1}\left( 
\frac{5}{2};3;\frac{-1}{4\alpha }\right) \int_{0}^{\infty }d\beta
\,_{1}F_{1}\left( \frac{3}{2};2;\frac{-1}{4\beta }\right) e^{-\left( \alpha
+\beta \right) Q^{2}\rho ^{2}}  \label{cinstqq}
\end{eqnarray}
(where $_{1}F_{1}\left( a;b;z\right) $ is the confluent hypergeometric
function \cite{abr65}) for the quark condensate coefficient and 
\begin{eqnarray}
C_{g\bar{q}\sigma Gq}^{\left( I+\bar{I}\right) }\left( Q^{2}\right) &=&-%
\frac{3\pi ^{2}}{2^{7}}\int d\rho n\left( \rho \right) \frac{\rho ^{6}}{\bar{%
m}_{q,1}\left( \rho \right) }  \nonumber \\
&&\times \int_{0}^{\infty }\frac{d\alpha }{\alpha ^{2}}\,_{1}F_{1}\left( 
\frac{5}{2};3;\frac{-1}{4\alpha }\right) \int_{0}^{\infty }\frac{d\beta }{%
\beta ^{2}}\,_{1}F_{1}\left( \frac{5}{2};3;\frac{-1}{4\beta }\right) \left(
\alpha +\beta \right) ^{2}e^{-\left( \alpha +\beta \right) Q^{2}\rho ^{2}}
\label{cinstqGq}
\end{eqnarray}
for the coefficient associated with the mixed quark-gluon operator $g_{s}%
\bar{q}\sigma Gq$. This operator appears for the first time in the OPE of
the pseudoscalar correlator. As expected, the above coefficients are not
suppressed by small quark masses and can be several orders of magnitude
larger than their perturbative counterparts. Since they arise from only one
quark propagating in the zero-mode state (while the other, soft one
contributes to the accompanying operator) we have denoted the corresponding
effective mass $\bar{m}_{q,1}\left( \rho \right) $.

Note that $\bar{m}_{q,1}$ does not equal $\bar{m}_{q,2}$ although both
emerge from a mean-field picture of quark interactions with soft vacuum
fields. The first main difference between the two is rooted in the fact that
they arise from averages over the ensemble of vacuum fields (approximated,
e.g., by instantons in the ILM). Due to the fluctuations in this ensemble
one should not expect averages over more than one zero-mode propagator to
factorize into separate averages over each propagator, and as a consequence $%
\bar{m}_{q,1}\neq \bar{m}_{q,2}$. This can be verified explicitly in the ILM
framework \cite{fac01}.

The second difference is specific to the IOPE: while in the two-zero-mode
contribution (\ref{cinst1}) the external momentum $Q$ is shared between both
quark lines, the {\it full} (i.e. maximal) $Q$ flows through the one
zero-mode quark line in (\ref{cinstqq}) and (\ref{cinstqGq}). Now, in more
complete treatments of the interactions with the vacuum background fields
the effective masses will become momentum-dependent quark self-energies $%
\bar{m}_{q}\left( k\right) $. For momenta much larger than the chiral
symmetry breaking scale, $k\gg \Lambda _{CSB}$, these self-energies become
insensitive to the soft CSB vacuum modes and approach the current quark
mass, $\bar{m}_{q}\left( k\rightarrow \infty \right) \rightarrow m_{q}$.
Since the effective mean-field masses can be considered as momentum averages
of such quark self-energies (or $\bar{m}_{q,i}=\bar{m}_{q}\left( \bar{Q}%
_{i}\right) $ with $\bar{Q}_{i}$ the typical momentum scale), one expects a
scale hierarchy 
\begin{equation}
\bar{m}_{q}\left( Q=0\right) \geq \bar{m}_{q,2}\geq \bar{m}_{q,1}\geq m_{q}
\label{mbounds}
\end{equation}
for all $\rho $. In order to get an idea of the size of $\bar{m}_{q,1}=\bar{m%
}_{q,1}\left( \bar{\rho},\bar{Q}_{1}\right) $, we will adopt the value of
the quark self-energy $M\left( k\right) $ in the large-$N_{c}$ approximation
to the ILM \cite{dia86} at the momentum transfer $\bar{Q}_{1}=1.5$ GeV (the
mean value of the interval $Q\in \left[ 1,2\right] $ GeV relevant for the
sum rule below)\footnote{%
The analogous estimate for $\bar{m}_{q,2}$ yields the value $\bar{m}%
_{q,2}\simeq 85$ MeV of Ref. \cite{fac01} at $\bar{Q}_{2}/\Lambda _{CSB}\sim
0.85$, which might explain why $\bar{m}_{q,2}$ is significantly smaller than
typical constituent quark masses.}, 
\begin{equation}
\bar{m}_{q,1}\sim M\left( \bar{Q}=1.5\text{ GeV}\right) =\frac{M\left(
0\right) \bar{Q}^{2}}{4\pi ^{2}\bar{\rho}^{2}}\varphi ^{2}\left( \bar{Q}%
\right) \simeq 20\text{ MeV,}
\end{equation}
where $M\left( 0\right) \simeq 0.3$ GeV and $\varphi \left( k\right) $ is a
combination of modified Bessel functions given in \cite{dia86}. Since $\bar{m%
}_{q,1}\simeq 0.02$ GeV is rather close to the lower end of the admissible
region (\ref{mbounds}), the size of the instanton-induced coefficients (\ref
{cinstqq}) and (\ref{cinstqGq}) will reach about one half of their upper
bound. The impact of different choices for $\bar{m}_{q,1}$ on predictions
and stability of the sum rule will be discussed in \cite{for202}.

Having calculated the IOPE of the pseudoscalar correlator up to $d=6$, we
now turn to the associated QCD sum rule. It will be convenient to rewrite
the different parts $\Pi ^{\left( X\right) }\left( Q^{2}\right) $ ($X\in
\left\{ pert,I+\bar{I},...\right\} $) of the correlator by means of the
dispersion relation as 
\begin{equation}
\Pi ^{\left( X\right) }\left( \tau \right) \equiv \hat{B}_{\tau }\Pi
^{\left( X\right) }\left( Q^{2}\right) =\frac{1}{\pi }\hat{B}_{\tau }\int ds%
\frac{%
\mathop{\rm Im}%
\Pi ^{\left( X\right) }\left( -s\right) }{s+Q^{2}}=\frac{1}{\pi }\int ds%
\mathop{\rm Im}%
\Pi ^{\left( X\right) }\left( -s\right) e^{-s\tau }.  \label{disp}
\end{equation}
In Eq. (\ref{disp}) we have already applied the obligatoy Borel transform $%
\hat{B}_{\tau }$ which improves IOPE convergence, removes subtraction terms
and emphasizes the ground-state contribution to the correlator \cite{shi79}.

A spectral sum rule is then obtained by equating the IOPE description $\Pi
^{\left( IOPE\right) }\left( \tau \right) $ of the correlator (in the $\tau $
region where it is reliable, see below) to a standard hadronic
representation $\Pi ^{\left( phen\right) }\left( \tau \right) $ whose
spectral function consists of pole and duality-continuum parts: 
\begin{equation}
\mathop{\rm Im}%
\Pi ^{\left( phen\right) }\left( -s;s_{0}\right) =%
\mathop{\rm Im}%
\Pi ^{\left( pole\right) }\left( -s\right) +%
\mathop{\rm Im}%
\Pi ^{\left( cont\right) }\left( -s;s_{0}\right) .
\end{equation}
The effective threshold $s_{0}$ delimits the duality interval of the
continuum into which we include, besides the standard OPE part, the
instanton contributions: 
\begin{equation}
\mathop{\rm Im}%
\Pi ^{\left( cont\right) }\left( -s;s_{0}\right) =\left[ 
\mathop{\rm Im}%
\Pi ^{\left( OPE\right) }\left( -s\right) +%
\mathop{\rm Im}%
\Pi ^{\left( I+\bar{I}\right) }\left( -s\right) \right] \theta \left(
s-s_{0}\right) .  \label{impicont}
\end{equation}
The instanton part will play an important r\^{o}le in the ensuing sum-rule
analysis. In the pole (i.e. resonance) contribution we allow, besides the
pion, also for its first excitation $\pi ^{\prime }$, 
\begin{equation}
\mathop{\rm Im}%
\Pi ^{\left( pole\right) }\left( s\right) =\pi \lambda _{\pi }^{2}\delta
\left( s-m_{\pi }^{2}\right) +\pi \lambda _{\pi ^{\prime }}^{2}\delta \left(
s-m_{\pi ^{\prime }}^{2}\right) ,
\end{equation}
where $m_{\pi }$ and $m_{\pi ^{\prime }}$ are the masses of the pion and the 
$\pi ^{\prime }$, and $\lambda _{\pi }=\sqrt{2}f_{\pi }K$ ($f_{\pi }^{\left(
\exp \text{t}\right) }=93\,$MeV) with $K=m_{\pi }^{2}/\left(
m_{u}+m_{d}\right) $. Including the $\pi ^{\prime }$ resonance explicitly
enables us to predict it's strength $\lambda _{\pi ^{\prime }}^{2}$ from the
sum-rule analysis. Thus we can directly determine the quantitative impact of
the $\pi ^{\prime }$ and decide whether it dominates the $\pi $ contribution
(as suggested in \cite{ste97}) or whether it can be absorbed into the
duality continuum (as in other QCD sum-rules).

Subtracting the continuum contributions of Eq. (\ref{impicont}) from the
IOPE, separately for each operator $\hat{O}_{n}$, we can write the sum rule
as 
\begin{eqnarray}
{\cal R}\left( \tau ;s_{0}\right) &\equiv &\Pi ^{\left( IOPE\right) }\left(
\tau \right) -\Pi ^{\left( cont\right) }\left( \tau ;s_{0}\right)  \nonumber
\\
&=&\sum_{n}\left[ {\cal R}_{\hat{O}_{n}}^{\left( pert\right) }\left( \tau
;s_{0}\right) +{\cal R}_{\hat{O}_{n}}^{\left( I+\bar{I}\right) }\left( \tau
;s_{0}\right) \right] =\lambda _{\pi }^{2}e^{-m_{\pi }^{2}\tau }+\lambda
_{\pi ^{\prime }}^{2}e^{-m_{\pi ^{\prime }}^{2}\tau },  \label{sr}
\end{eqnarray}
where the pole contributions are isolated on the right-hand-side and where
we have defined 
\begin{eqnarray}
{\cal R}_{\hat{O}_{n}}^{\left( X\right) }\left( \tau ;s_{0}\right) &\equiv
&\Pi _{\hat{O}_{n}}^{\left( X\right) }\left( \tau \right) -\frac{1}{\pi }%
\int_{0}^{\infty }ds%
\mathop{\rm Im}%
\Pi _{\hat{O}_{n}}^{\left( X\right) }\left( -s\right) \theta \left(
s-s_{0}\right) e^{-s\tau }  \nonumber \\
&=&\frac{1}{\pi }\int_{0}^{s_{0}}ds\left\langle \hat{O}_{n}\right\rangle 
\mathop{\rm Im}%
C_{\hat{O}_{n}}^{\left( X\right) }\left( -s\right) e^{-s\tau }.  \label{r}
\end{eqnarray}

It remains to calculate the imaginary parts of the Wilson coefficients in
the timelike region from the explicit expressions for the $C_{\hat{O}%
}^{\left( X\right) }$ given above. For the perturbative Wilson coefficients
we find 
\begin{eqnarray}
\mathop{\rm Im}%
C_{1}^{\left( pert\right) }\left( -s\right) &=&\frac{3}{8\pi }s\left\{ 1+%
\frac{\alpha _{s}}{\pi }\left[ \frac{17}{3}-2\ln \left( \frac{s}{\mu ^{2}}%
\right) \right] \right\} , \\
\mathop{\rm Im}%
C_{\bar{q}q}^{\left( pert\right) }\left( -s\right) &=&-\pi m_{q}\delta
\left( s\right) ,\text{ \ \ \ \ \ }%
\mathop{\rm Im}%
C_{\alpha G^{2}}^{\left( pert\right) }\left( -s\right) =\frac{1}{8}\delta
\left( s\right) ,\text{ \ \ \ \ \ } \\
\mathop{\rm Im}%
C_{O_{1}}^{\left( pert\right) }\left( -s\right) &=&-%
\mathop{\rm Im}%
C_{O_{2}}^{\left( pert\right) }\left( -s\right) =\pi ^{2}\delta ^{\prime
}\left( s\right) .
\end{eqnarray}

The instanton-induced contributions can be obtained in closed form from the
imaginary parts of the coefficients (\ref{cinst1}), (\ref{cinstqq}) and (\ref
{cinstqGq}). The unit-operator coefficient (\ref{cinst1}) gives 
\begin{equation}
\mathop{\rm Im}%
C_{1}^{\left( I+\bar{I}\right) }\left( -s\right) =-\frac{\pi ^{2}}{2}\int
d\rho n\left( \rho \right) \rho ^{2}\frac{\bar{m}_{u,2}^{2}\left( \rho
\right) +\bar{m}_{d,2}^{2}\left( \rho \right) }{\bar{m}_{u,2}^{2}\left( \rho
\right) \bar{m}_{d,2}^{2}\left( \rho \right) }sJ_{1}\left( \sqrt{s}\rho
\right) Y_{1}\left( \sqrt{s}\rho \right) .
\end{equation}
The imaginary parts of the coefficients associated with
chiral-symmetry-breaking operators have a more complex structure. They can
be expressed in terms of the integrals 
\begin{equation}
I_{Ji,j}\left( s\right) =\int_{0}^{1}d\eta \frac{\eta ^{j+3}J_{i}\left( 
\sqrt{s}\rho \eta \right) }{\sqrt{1-\eta ^{2}}},\,\text{\ \ \ \ \ \ \ \ }%
I_{Yi,j}\left( s\right) =\int_{0}^{1}d\eta \frac{\eta ^{j+3}Y_{i}\left( 
\sqrt{s}\rho \eta \right) }{\sqrt{1-\eta ^{2}}},
\end{equation}
where $J_{i}\left( z\right) $ and $Y_{i}\left( z\right) $ are Bessel and
Neumann functions \cite{abr65}. For the instanton contribution to the
quark-condensate coefficient (\ref{cinstqq}) we find 
\begin{equation}
\mathop{\rm Im}%
C_{\bar{q}q}^{\left( I+\bar{I}\right) }\left( -s\right) =-\frac{2^{6}\pi ^{2}%
}{3}\int d\rho \frac{n\left( \rho \right) \rho ^{4}}{\bar{m}_{q,1}\left(
\rho \right) }I_{J1,0}\left( s\right) I_{Y1,0}\left( s\right) -\frac{%
2^{3}\pi ^{3}}{3}\int d\rho \frac{n\left( \rho \right) \rho ^{2}}{\bar{m}%
_{q,1}\left( \rho \right) }\delta \left( s\right)  \label{imcqq}
\end{equation}
and for the mixed condensate coefficient (\ref{cinstqGq}) 
\begin{eqnarray}
\mathop{\rm Im}%
C_{g\bar{q}G\sigma q}^{\left( I+\bar{I}\right) }\left( -s\right) &=&\frac{%
2^{2}\pi ^{2}}{3}\int d\rho \frac{n\left( \rho \right) \rho ^{6}}{\bar{m}%
_{q,1}\left( \rho \right) }\left[ 2I_{J0,1}\left( s\right) I_{Y0,1}\left(
s\right) -I_{J1,2}\left( s\right) I_{Y1,0}\left( s\right) \right.  \nonumber
\\
&&-\left. I_{J1,0}\left( s\right) I_{Y1,2}\left( s\right) \right] -\frac{\pi
^{3}}{2^{2}}\int d\rho \frac{n\left( \rho \right) \rho ^{4}}{\bar{m}%
_{q,1}\left( \rho \right) }\delta \left( s\right) .  \label{imcqGq}
\end{eqnarray}
Note that both (\ref{imcqq}) and (\ref{imcqGq}) receive strong contributions
from $s=0$ which significantly reduce the $s_{0}$-dependence of the CSB
terms ${\cal R}_{\bar{q}q}\left( \tau ;s_{0}\right) $ and ${\cal R}_{g\bar{q}%
G\sigma q}\left( \tau ;s_{0}\right) $ (cf. Fig. 1b below).

For the quantitative analysis of the sum rule (\ref{sr}) we fix the IOPE
parameters at standard values: $\Lambda _{QCD}=0.2$ GeV, $\left\langle \bar{q%
}q\right\rangle =-0.0156\,{\rm GeV}^{3}$, $\langle \alpha _{s}G^{2}\rangle
=0.04\,{\rm GeV}^{4}$, $\left\langle g_{s}\bar{q}\sigma Gq\right\rangle
=m_{0}^{2}\langle \bar{q}q\rangle $, $m_{0}^{2}=0.8$ GeV$^{2}$, $%
\left\langle \hat{O}_{4}+\hat{O}_{5}\right\rangle =\left( 56\pi \alpha
_{s}/27\right) \left\langle \bar{q}q\right\rangle ^{2}$. As in previous IOPE
sum rules, we approximate the instanton distribution as $n\left( \rho
\right) =\bar{n}\delta \left( \rho -\bar{\rho}\right) $ with\footnote{%
Phenomenological, ILM and lattice evidence for these scales is discussed in 
\cite{sch98}.} $\bar{n}=0.5$ fm$^{-4}$ and $\bar{\rho}=0.33$ fm. The
standard RG improvement, finally, amounts to scaling $\left\langle \hat{O}%
_{n}\right\rangle \rightarrow \xi ^{2-\gamma _{n}}\left\langle \hat{O}%
_{n}\right\rangle $ with $\gamma _{1,\alpha _{s}G^{2}}=0$, $\gamma _{\bar{q}%
q}=1$, $\gamma _{g_{s}\bar{q}\sigma Gq}=-1/6$ and 
\begin{equation}
\xi \left( \tau \right) =-\frac{1}{2}\ln \left( \tau \Lambda ^{2}\right)
^{-4/9}\left[ 1-\frac{290}{729}\frac{1}{\ln \left( \tau \Lambda ^{2}\right) }%
+\frac{256}{729}\frac{\ln \left[ -\ln \left( \tau \Lambda ^{2}\right) \right]
}{\ln \left( \tau \Lambda ^{2}\right) }\right] ,
\end{equation}
to replacing $\alpha _{s}$ by the (two-loop) running coupling, and to
substituting $\mu ^{2}\rightarrow 1/\tau $.

The resulting $\tau $- and $s_{0}$-dependence of ${\cal R}\left( \tau
;s_{0}\right) $ is plotted in Fig. 1a. Note that for $\tau \gtrsim 1.4$ GeV$%
^{-2}$ and $s_{0}\gtrsim 2.5$ GeV$^{2}$, ${\cal R}\left( \tau ;s_{0}\right) $
becomes practically $s_{0}$-independent. This is a consequence of the
exponential $\exp \left( -s\tau \right) $ in the integrands of (\ref{r})
which renders the dispersive integrals insensitive to their upper
integration limit $s_{0}$ whenever $\exp \left( -s_{0}\tau \right) \lesssim
10^{-2}$. Therefore, $s_{0}$ cannot be determined by the sum rule in this $%
\left( \tau ,s_{0}\right) $-region. Another conspicuous feature of ${\cal R}%
\left( \tau ;s_{0}\right) $ is the valley at intermediate $s_{0}$ and small $%
\tau $, which can be traced to the instanton continuum contribution of the
unit-operator coefficient. In this $\left( \tau ,s_{0}\right) $-region the
sum rule does not match since the positive slopes in $\tau $-direction do
not fit the decaying exponentials of the pole contributions. In the $s_{0}$%
-region around 2 GeV$^{2}$ (i.e. around the values which one would expect
from duality arguments), however, ${\cal R}\left( \tau ;s_{0}\right) $ shows
an extended ``mountain ridge'' with the slow decay in $\tau $ which matches
an exponential containing a rather small mass. Thus we have qualitatively
identified the area in which the sum rule will optimize, and we have found
that the pion contribution must be sizeable. Both observations will be
confirmed by the quantitative analysis below.

Figure 1b exhibits the $\tau $- and $s_{0}$-dependence of the new,
instanton-induced CSB contributions to ${\cal R}\left( \tau ;s_{0}\right) $,
i.e. of 
\begin{equation}
{\cal R}^{\left( CSB\right) }\left( \tau ;s_{0}\right) \equiv {\cal R}_{\bar{%
q}q}^{\left( I+\bar{I}\right) }\left( \tau ;s_{0}\right) +{\cal R}_{g\bar{q}%
G\sigma q}^{\left( I+\bar{I}\right) }\left( \tau ;s_{0}\right) .
\end{equation}
The essential message of this plot is that in the physically meaningful $%
\left( \tau ,s_{0}\right) $-region where $s_{0}\leq 4$ GeV$^{2}$ or $\tau
\geq 0.4$ GeV$^{-2}$, the CSB contributions are monotonically {\it increasing%
} with $\tau $. In fact, those are the {\it only} relevant contributions to
the IOPE whose slope is positive. Moreover, the slope at small $\tau $
becomes maximal in the $s_{0}$-region around 2 GeV$^{2}$ where the sum rule
optimally matches. Thus these CSB contributions overcome much of the
negative slope originating from the chirally invariant operators (with both
perturbative and instanton-induced coefficients) and thereby lower the
prediction for the pion mass. This qualitative result demonstrates how the
CSB condensates, ``activated'' by direct instantons, can reconcile the
pseudoscalar sum rule with Goldstone's theorem. As a byproduct, they enhance
the strength of the pion contributions relative to those from higher
resonances.

The main task of the quantitative sum-rule analysis is to find those values
of the hadronic parameters to be determined for which both sides of (\ref{sr}%
) optimally match in the fiducial $\tau $-region. The latter is obtained
from the standard requirements that (i) the highest-dimensional operators
should contribute less than 10\% to ${\cal R}\left( \tau ;s_{0}\right) ,$
that (ii) multi-instanton effects should be negligible, and that (iii) the
continuum contributions should be limited to maximally 50\% of the total $%
{\cal R}\left( \tau ;s_{0}\right) $. Since one sum rule cannot reliably
determine all five hadronic parameters, we have chosen to fix those best
known from experiment, $m_{\pi ^{0}}=135$ MeV and $m_{\pi ^{\prime }}=1300$
MeV \cite{gro00}, and to determine the values of the remaining ones\footnote{%
Alternative analysis strategies will be considered in \cite{for202}. We have
checked, in particular, that the sum rule predicts the small pion mass
scale, $m_{\pi }\simeq 140$ MeV, if $\lambda _{\pi }$ is fixed at it's
average phenomenological value $\lambda _{\pi }\simeq 0.27$ GeV$^{2}$.},
i.e. $s_{0}$, $\lambda _{\pi }$ and $\lambda _{\pi ^{\prime }}$. Both sides
of the optimized sum rule are shown in Fig. 2. Their good match confirms the
consistency and stability of the sum rule. The resulting values for the
couplings are $\lambda _{\pi }^{2}=0.078$ GeV$^{4}$ and $\lambda _{\pi
^{\prime }}^{2}=0.032$ GeV$^{4}$, while the continuum threshold becomes $%
s_{0}=1.84$ GeV$^{2}$. With $f_{\pi }=93\,$MeV and $m_{\pi }=135$ MeV this
implies $m_{u}+m_{d}=\sqrt{2}f_{\pi }m_{\pi }^{2}/\lambda _{\pi }\simeq 9.0$
MeV (at $\mu \sim 1$ GeV), within the estimated range ($6.75-16.2$ MeV) of
Ref. \cite{gro00}.

Fig. 2 also shows the individual contributions ${\cal R}_{1+\alpha
G^{2}+O_{4}+O_{5}}^{\left( pert\right) }$, ${\cal R}_{1}^{\left( I+\bar{I}%
\right) }$, ${\cal R}_{\bar{q}q}^{\left( I+\bar{I}\right) }$, and ${\cal R}%
_{g\bar{q}G\sigma q}^{\left( I+\bar{I}\right) }$ to the left-hand side of
the sum rule. The instanton contributions can be seen to dominate, and the
positive slope of the CSB parts indeed compensates most of the negative
slope brought in by the chirally invariant ones. As expected, ${\cal R}%
_{1}^{\left( I+\bar{I}\right) }$ contributes with (strongly) negative slope.
This explains why the sum rule of Ref. \cite{ste97}, which took exclusively
this direct-instanton contribution (without it's continuum part) into
account, could not be stabilized without a strong resonance in the 1 GeV
region.

Implementing the CSB operators (and their continuum contributions) increases
the relative strength of the pion pole to $\left( \lambda _{\pi
}^{2}/\lambda _{\pi ^{\prime }}^{2}\right) \exp \left( m_{\pi ^{\prime
}}-m_{\pi }\right) \simeq 8-70$ in the fiducial Borel domain. As a
consequence, the pion dominates while the $\pi ^{\prime }$ contributions
become negligible. In conjunction with the relatively small value of the
continuum threshold ($s_{0}=1.84$ GeV$^{2}$) this suggests that the
excited-state contributions can be absorbed into the duality continuum (as
in practically all other sum-rule channels). The pion dominance seems
natural in view of the exceptionally large mass difference between ground
state and first excitation in the pseudoscalar channel. Moreover, it is
consistent with lattice simulations of mesonic point-to-point correlators
which find the pseudoscalar correlator well described by just the ground
state pole and the duality continuum \cite{chu93}.

In summary, we have introduced instanton-generated Wilson coefficients
associated with the leading chiral-symmetry-breaking operators in the IOPE
of the pseudoscalar correlator. The new contributions are fully
nonperturbative (both at soft and hard momenta) and supply previously
missing information about spontaneous chiral symmetry breaking which
reconciles the associated pseudoscalar sum rule with Goldstone's theorem. As
a consequence, this sum rule becomes the first in its channel which is able
to reproduce the light mass scale of the pion. This resolves the puzzle
stated in the introduction. Moreover, the chirally-odd operators suppress
the contributions from higher-lying resonances, which can therefore be
subsumed into the dispersive continuum. Both effects, as well as the
stability of the sum rule, are enhanced by the instanton-induced continuum
contributions. Additional implications of the new operators, e.g. for the
calculation of the light quark mass values on the basis of pseudoscalar sum
rules, will be discussed elsewhere \cite{for202}. \

\newpage

\section{Figure captions}

\begin{enumerate}
\item  Fig. 1a: The theoretical side of the sum rule, ${\cal R}\left( \tau
;s_{0}\right) $, with $\bar{m}_{q,1}=20$ MeV and $\bar{m}_{q,2}=85$ MeV.

\item  Fig. 1b: The instanton-induced contributions of the
chiral-symmetry-breaking operators, ${\cal R}^{\left( CSB\right) }\left(
\tau ;s_{0}\right) \equiv {\cal R}_{\bar{q}q}^{\left( I+\bar{I}\right)
}\left( \tau ;s_{0}\right) +{\cal R}_{g\bar{q}G\sigma q}^{\left( I+\bar{I}%
\right) }\left( \tau ;s_{0}\right) $, to the theoretical side of the sum
rule.

\item  Fig. 2: The right-hand-side (full line) of the optimized sum rule is
compared to the theoretical side ${\cal R}\left( \tau ;s_{0}=1.84\text{ GeV}%
\right) $ (dotted). The contributions to ${\cal R}$ from the perturbative ($%
{\cal R}^{\left( pert\right) },$ short-dashed) and instanton-induced (${\cal %
R}_{1}^{\left( I+\bar{I}\right) }$ dot-dashed, ${\cal R}_{\bar{q}q}^{\left(
I+\bar{I}\right) }$ dot-dot-dashed, ${\cal R}_{g\bar{q}G\sigma q}^{\left( I+%
\bar{I}\right) }$ short-dotted, their sum dashed) Wilson coefficients are
also plotted separately.
\end{enumerate}

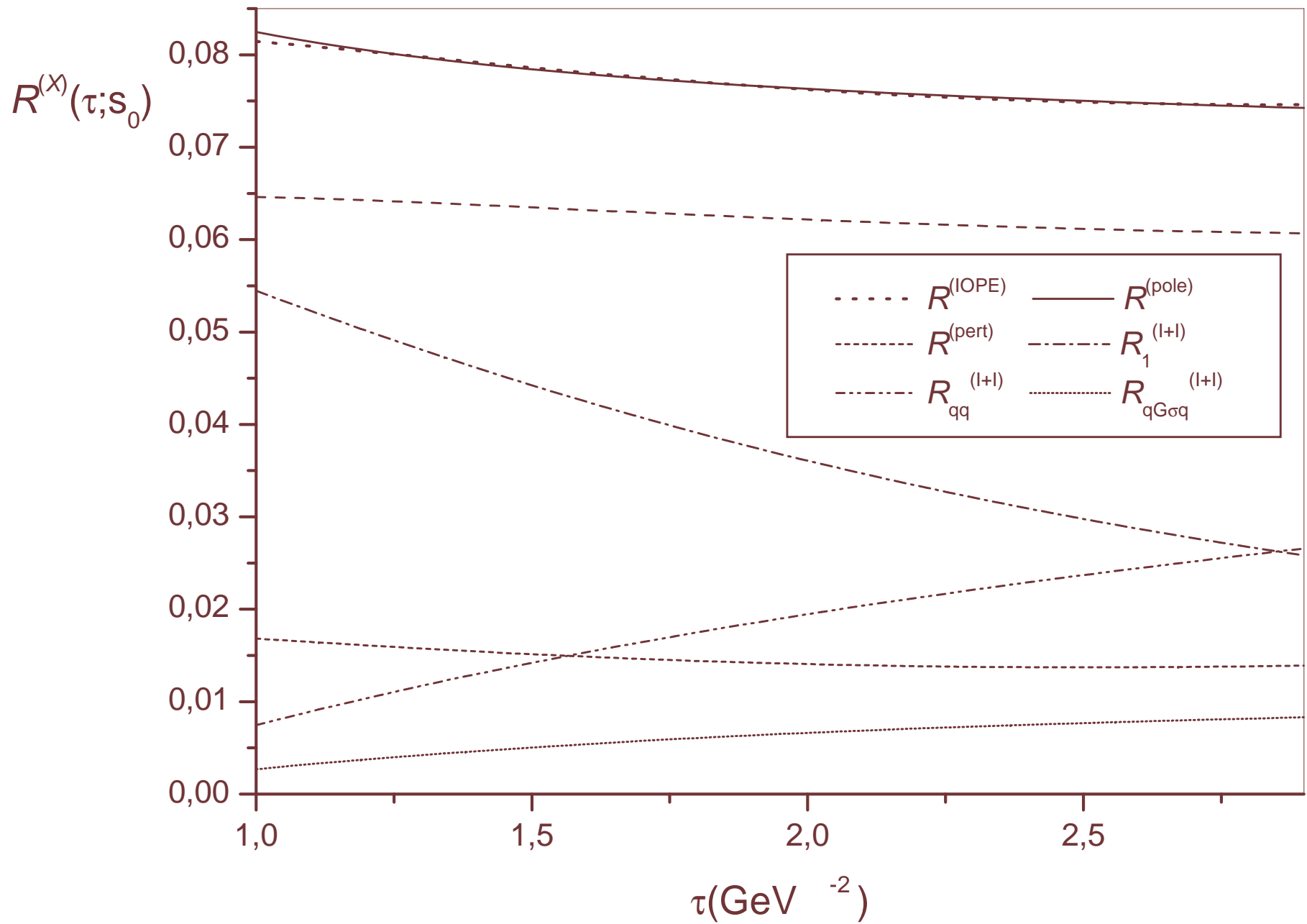
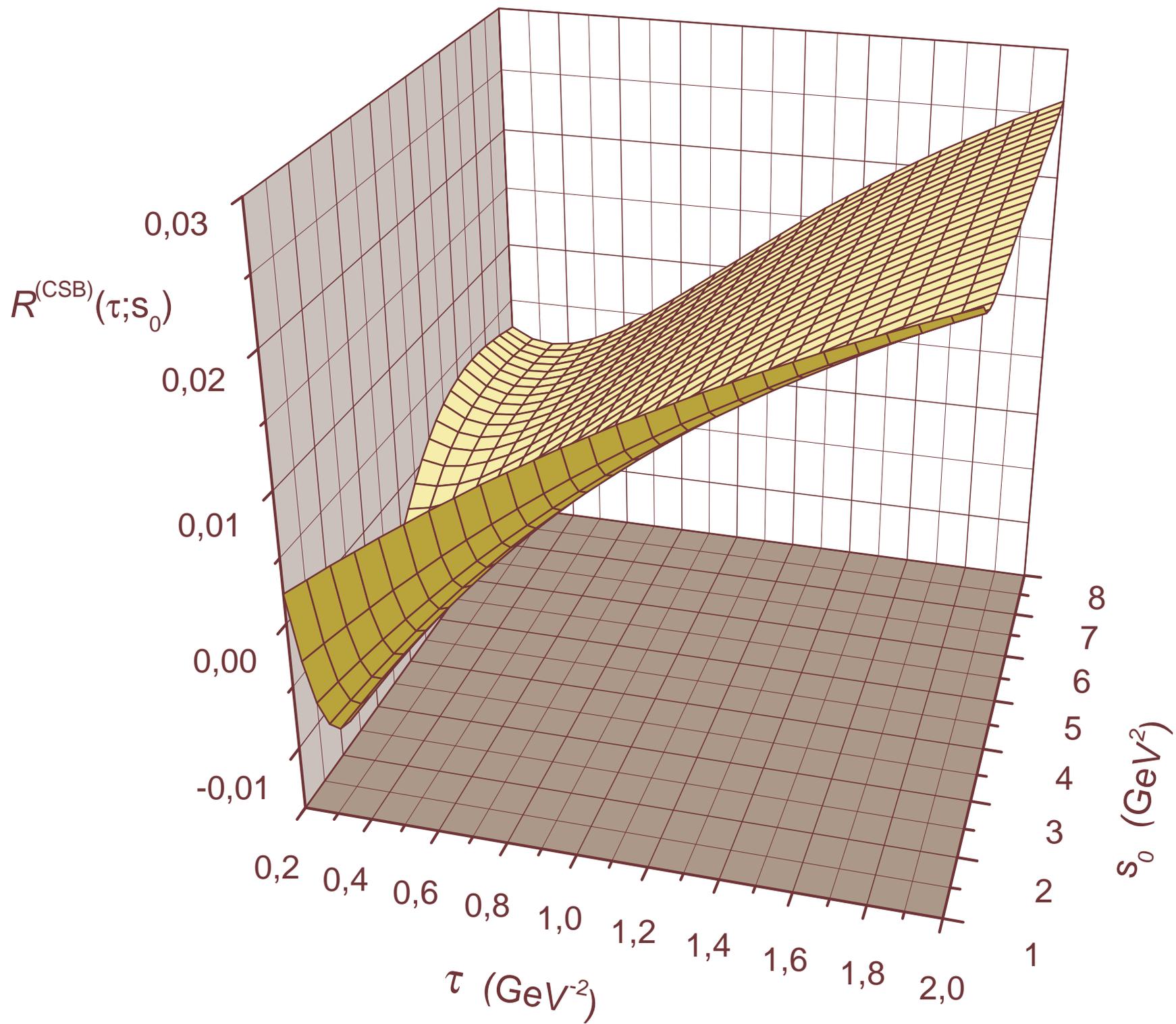
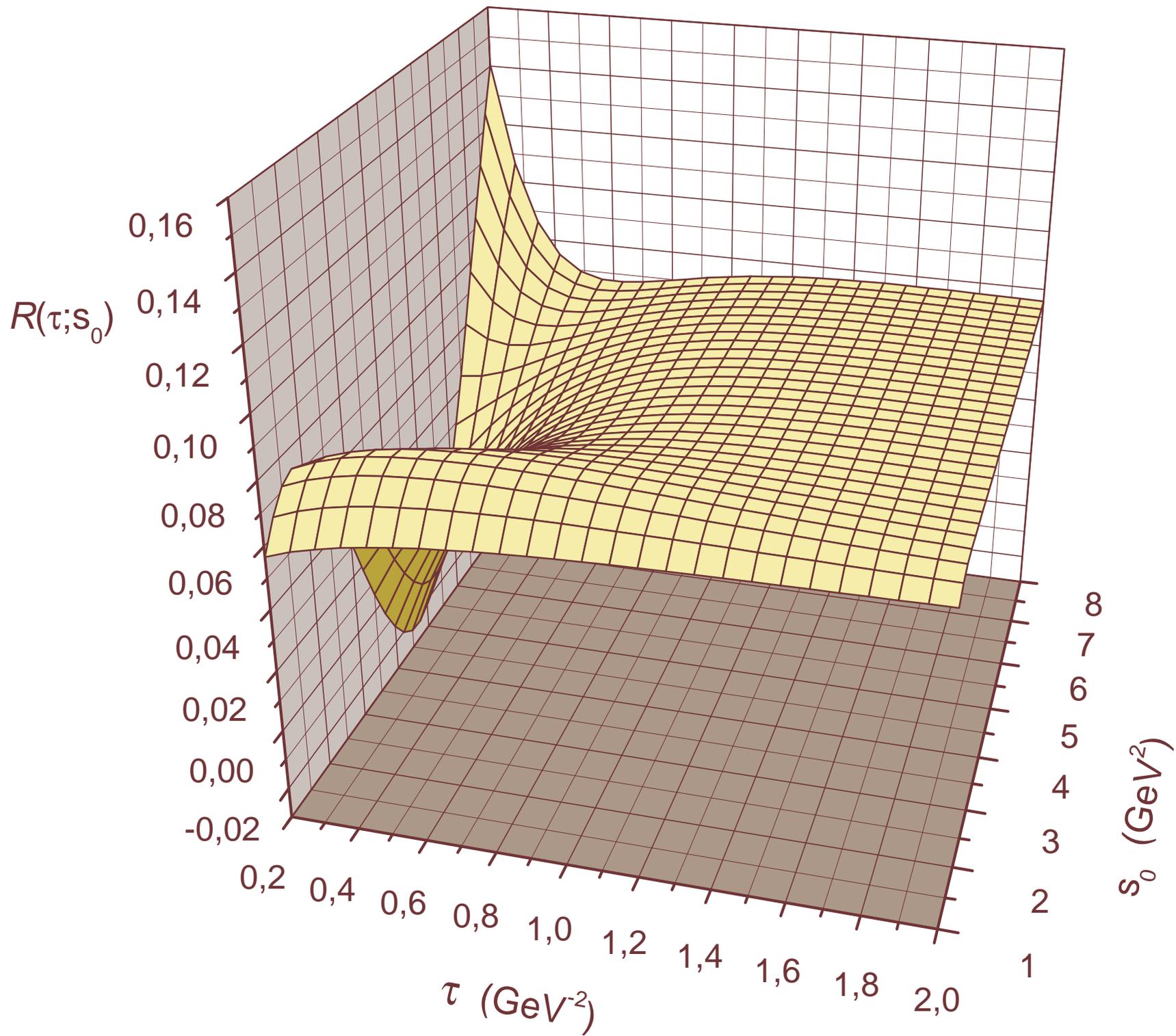
\end{document}